# Strain gradient drives lithium dendrite growth from the atomic-scale simulations

Gao Xu,[1] Feng Hao[2], Jiawang Hong,[3,*] and Daining Fang[1,4]

[1]*State Key Laboratory for Turbulence and Complex Systems & Center for Applied Physics and Technology, College of Engineering, Peking University, Beijing 100871, PR China*

[2]*Department of Engineering Mechanics, Shandong University, Jinan 250100, China*

[3]*School of Aerospace Engineering, Beijing Institute of Technology, Beijing 100081, China*

[4]*Institute of Advanced Structure Technology, Beijing Institute of Technology, Beijing 100081, PR China*

[*]Corresponding author

*E-mail address*: hongjw@bit.edu.cn

**Abstract:**

Dendrite formation is a major obstacle, such as capacity loss and short circuit, to the next-generation high-energy-density lithium (Li) metal batteries. The development of successful Li dendrite mitigation strategies is impeded by an insufficient understanding of Li dendrite growth mechanisms. Li-plating-induced internal stress in Li metal and its effect on dendrite growth have been studied in previous models and experiments, while the underlying microcosmic mechanism is elusive. Here, we analyze the role of plating-induced stress in dendrite formation through first-principles calculations and *ab initio* molecular dynamics simulations. We show that the deposited Li forms a stable atomic nanofilm structure on copper (Cu) substrate. It is found that the adsorption energy of Li atoms increases from the Li-Cu interface to deposited Li surface, leading to more aggregated Li atoms at the interface. Compared to the pristine Li metal, the deposited Li in the early stage becomes compacted and suffers in-plane compressive stress. Interestingly, we find that there is a giant strain gradient distribution from the Li-Cu interface to deposited Li surface, which makes the deposited atoms adjacent to the Cu surface tend to press upwards with perturbation, causing the dendrite growth. This understanding provides an insight to the atomic-scale origin of Li dendrite growth and may be useful for suppressing the Li dendrite in the Li-metal-based rechargeable batteries.

Project supported by the National Key R&D Program of China (2016YFB0700600), the Foundation for Innovative Research Groups of the National Natural Science Foundation of China (No. 11521202, No. 11572040 and No. 11804023), the Thousand Young Talents Program of China, National Key Research and Development Program of China (Grant No. 2016YFB0402700) are gratefully acknowledged.

# 1 Introduction

Li metal offers the highest theoretical capacity (3,860 mAh g$^{-1}$) and the lowest electrochemical potential (−3.04 V versus standard hydrogen electrode) amongst all anode materials[1]. In recent years, Li-metal-based rechargeable batteries, including Li–sulfur, Li–air and Li–selenium batteries, are actively being developed for electric vehicles and grid-scale storage because of their much higher energy densities compared with Li-ion batteries[2-3]. However, the practical application of Li metal anodes is critically impeded by the formation of Li dendrites[4]. The uncontrollable dendrite growth continuously consumes the active materials and potentially punctures the separator, resulting in irreversible capacity loss, short circuits, and even safety hazards[2,5-6]. Much effort has been devoted to suppressing dendrite growth[7-10], such as using liquid electrolyte additives[1,11-12] or solid electrolytes[13-14], application of mechanical pressure and modification of substrate smoothness[15], adoption of different charging methods[16], artificial solid-electrolyte interface[17-18], structural design of the electrodes and current collectors[19-22].

For the aforementioned efforts to tackle the Li dendrite problem, stress/strain is one of the most critical and fundamental aspects. For example, Zhang et al.[6] addressed the problem of dendrite suppression by applying two-dimensional materials to Li metal systems and preventing dendrite penetration through stress release and mechanical blocking. He et al.[23] recently reported that Li whiskers can yield, buckle, kink or stop growing under certain elastic constraints. In addition, the presence of plating-induced stress in Li metal has an impact on Li growth[8]. For instance, interfacial stress generated at the solid-solid interfaces between the electrode and the solid electrolyte and its role in the interfacial stability and dendrite initiation were studied[24-25]. Many microstructural evolution phenomena in materials are driven by stress. For example, whiskers can grow from tin films under compressive stress[26]. Since residual stress is omnipresent in the metal plating process[27-28], stress would exist during Li electrodeposition, which may play an important role in the filamentary Li dendrite growth.

In fact, Li dendrite growth as a stress relief mechanism has been proposed[29], while a detailed working mechanism and experimental support are lacking. Wang et al.[8] identified the presence of compressive stress during plating and its role in promoting Li dendrite initiation. They also revealed that Li dendrites can be effectively suppressed in carbonate-based electrolytes by relieving the residual stress using soft substrates. They proposed a stress-driven dendrite growth mechanism to explain the experimental observations, and a critical stress level was predicted as an important criterion for achieving dendrite-free Li electroplating. Their development of compressive stress during thin film growth is common for materials that have high atomic mobility and grow according to the Volmer-Weber mechanism[30-31]. The origin of the in-plane film compression is attributed to the insertion of excessive atoms into the grain boundary under the non-equilibrium growth conditions[32]. The previous studies on Li dendrite growth mechanisms are mainly from a macroscale or mesoscale perspective. However, the Li dendrite growth is also a nanoscale issue, such as the nucleation and growth behavior, and the corresponding internal stress needs to be explored at the atomic-scale.

In this work, the microscopic mechanism of Li dendrite growth is investigated. We analyze the formation of plating-induced stress in deposited Li film on Cu substrate by using the first-principles calculations and molecular dynamics simulations. We find the adsorption energy of Li atoms in the Li-Cu interface is lower than those located near deposited surface, which makes deposited Li atoms tend to occupy the position close to the Li-Cu interface. This causes the region near the Li-Cu interface holding more Li atoms with smaller atomic distance, which induces internal compressive stress. In addition, there forms a giant strain gradient distribution from Li-Cu interface to the deposited Li surface during Li deposition, which may drive the dendrite growth. This atomic investigation may help understanding the Li dendrite growth mechanism and shed light on the extensive pursuit of Li dendrite suppression strategies with potential implications for other metallic electrode systems.

## 2 Calculation Details

The calculations are performed using Vienna ab initio simulation package (VASP) [33-34], which is based on the density functional theory (DFT). Projector-augmented-wave (PAW) potentials[35] are used to take into account the electron−ion interactions, while the electron exchange-correlation interactions are treated using generalized gradient approximation (GGA)[36] in the scheme of Perdew-Burke-Ernzerhof. A plane wave cutoff of 600 eV is set in our calculations. Starting from the obtained bulk Cu structures, we create slabs of Cu with 2 × 2 × 3 supercell and adding a ∼ 20 Å vacuum in the cell to simulate the Cu substrate, as shown in Fig. 1(a). For the following Li deposition calculations, the $z$-axis will be elongated to satisfy the 20 Å vacuum layer condition. The slab structure has two different layers of Cu atoms, labeled A and B in Fig. 1(b). There are six layers of Cu atoms in the cell and the bottom three layers are fixed during the relaxation. Every layer of Cu has eight atoms with the A type topmost layer, which is adjacent to deposited Li atoms. K-point samplings of 6 × 6 × 2 are used and the Γ point is included. Atomic relaxation is performed until the total energy converged to $1 \times 10^{-6}$ eV and the force on each atom is smaller than 0.01 eV/Å.

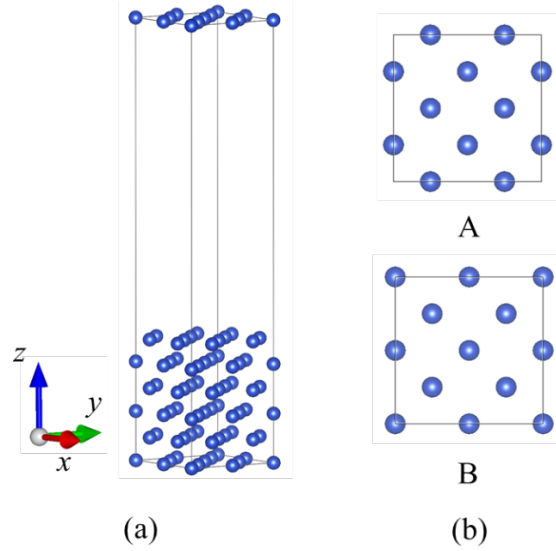

**Fig. 1** (a) The Cu slab calculation structure and (b) its two different layers of Cu atoms. The black frame indicates the cell of the calculation structure.

The kinetic properties of Li deposition on Cu are studied using the *ab initio* molecular dynamics (AIMD) method. The simulations are performed on the 4 × 4 × 3 supercell based on above-mentioned slab cell, containing 192 Cu atoms in total, and only the Γ point is used for the Brillouin zone sampling. We perform constant volume and temperature (NVT) AIMD calculations with a Nose thermostat[37] at 300 K for 3000 steps with a time step of 1 fs to investigate Li deposition on Cu substrate.

## 3 Results and Discussion

### 3.1 Ground states of initial deposited Li atoms on Cu substrate

Our calculated lattice constants for Li and Cu metal are 3.43 Å and 3.62 Å, respectively, which are in good agreement with the experimental values of 3.51 Å[38] and 3.60 Å[39]. The crystal structures of Li and Cu metal are body-centered cubic and face-centered cubic, respectively. To analyze the adsorption of Li on Cu substrate, we consider several configurations by adding Li atoms to the different positions on the Cu slab structure. The adsorption energy is defined as

$$E_a = (E_{ad-slab} - nE_{ad} - E_{slab})/n$$

where $E_{ad\text{-}slab}$ is the total energy of the compound in which Li is adsorbed on the Cu slab, $E_{ad}$ is the energy of an isolated Li atom, $E_{slab}$ is the energy of the relaxed Cu slab, and *n* is the number of adsorbed Li atoms. The geometry structure and adsorption energy are obtained after the positions are relaxed. It is worth noting that the dynamics effect is neglected for the calculations of this part, such as diffusion. The adsorption energy will be considered as the criterion to explore the most stable configuration.

To explore the configuration of initial deposited Li atoms on Cu substrate, we first consider two sites with high symmetry: the site on the top of a Cu atom (Top) and the site above the center of a square (Center), as shown in Fig. 2(a) and 2(b), respectively. By comparing adsorption energy, the center site is more stable than the top site, which means that the initial deposited Li atoms tend to adsorb on the center sites of Cu substrate surface. The distance between the Li atom at the center cite and the adjacent Cu atom is 2.60 Å. In contrast, the distances of two neighboring Li atoms and two neighboring Cu atoms are 2.97 Å ((111) orientation of Li metal) and 2.56 Å ((110) orientation of Cu metal), respectively. It indicates that the atomic spacing at the Li-Cu interface is between those of pristine Li and Cu metal, thus lattice mismatch occurs.

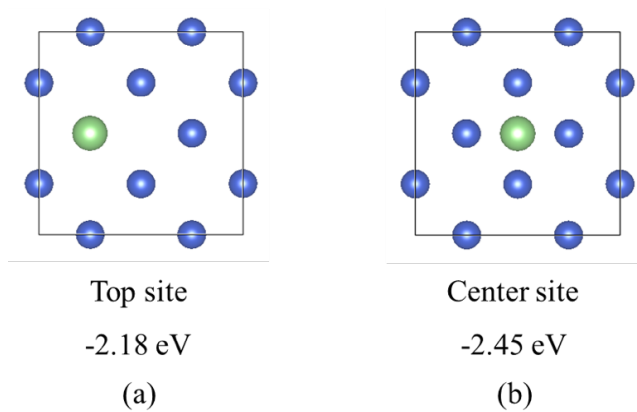

Top site     Center site
-2.18 eV     -2.45 eV
(a)     (b)

**Fig. 2** The top view of calculation cell and corresponding adsorption energies for the single Li atom adsorbed at (a) top site and (b) center cite. The green and blue balls represent Li and Cu atoms, respectively. Only the topmost Cu atom layer is shown for identifying structure clearly.

As Li atoms further deposit, more Li atoms will be adsorbed on Cu surface and form a deposited Li film. We construct some cells containing more than one Li atom on Cu slab surface, and then relax these structures and obtain their corresponding adsorption energies, as shown in Fig. 3. For each configuration, the relaxed Li atom position has the same *z*-coordinate, representing one single Li atomic layer. For one Li atomic layer adsorbed on Cu surface, the most stable configuration is that four Li atoms are adsorbed on center sites in the cell, as shown in Fig. 3(b); meanwhile, the in-plane spacing between Li atoms is the same as that of Cu substrate atoms. The result suggests that the deposition and initial crystal growth of Li near the Cu surface may follow the Cu lattice parameter and Li crystal structure. Moreover, the subsequent deposited Li atoms tend to be closer to each other than those in pristine Li metal, causing the compressive stress in the deposited Li layers.

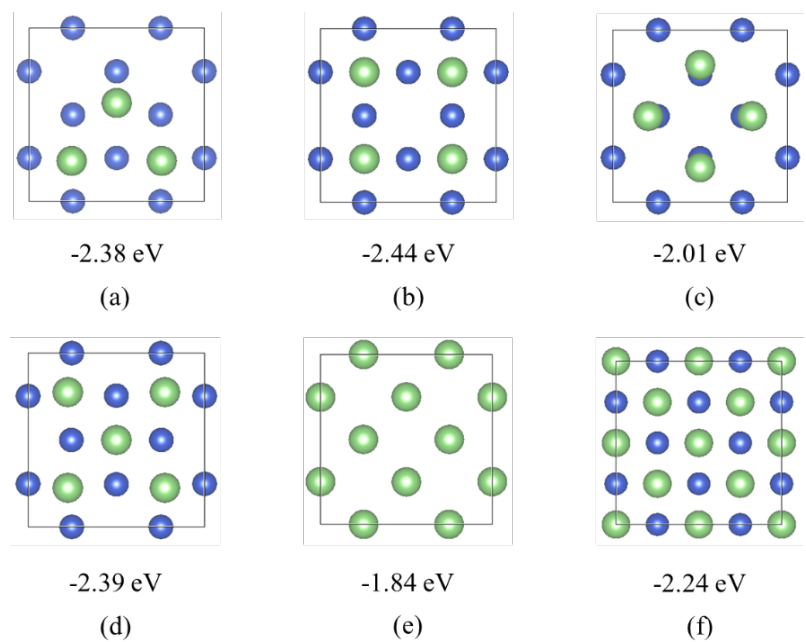

**Fig. 3** The top view of calculation cell and corresponding adsorption energies for the adsorbed Li atoms at different positions. The figure labeled (e) represents every Li atom is above the topmost layer Cu atoms. The energy showed in each pattern is the adsorption energy per Li atom.

In Fig. 4(a) the initial configurations, with the number of Li atomic layers from one to five, are chosen to obtain their adsorption energies. Fig. 4(b) shows the relationship between the adsorption energy and the number of Li atomic layers of initial configurations. As the number of Li layers increases, the adsorption energy becomes higher. It indicates that the Li atoms closer to the Li-Cu interface are more favored than those located near the deposited surface. Besides, for the initial configurations with two and three Li atomic layers, the Li atoms significantly descend closer to Cu surface after relaxation. Therefore, thanks to the lower adsorption energy, the space near the Cu surface tend to accommodate more deposited Li atoms.

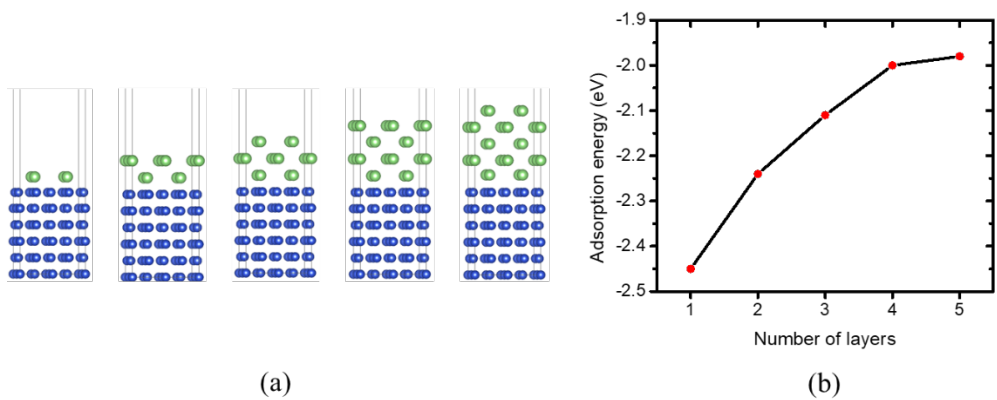

**Fig. 4** (a) The five initial calculation configurations with different numbers of Li atomic layers from one to five. (b) The adsorption energy as a function of the number of Li atomic layers of initial configurations.

**3.2 The dynamic simulations of deposited Li film on Cu substrate**

In previous section, we investigate the Li deposition on Cu substrate from static simulation. In this section, we will study the Li deposition behavior from *ab initio* molecular dynamic simulation. Based on the configuration of five Li atomic layers in Fig. 4(a), we create a $4 \times 4 \times 3$ supercell and add more Li atoms to simulate the configuration of deposited Li film on Cu substrate at room-temperature, the initial configuration is shown in Fig. 5(a) (192 Cu atoms and 160 Li atoms). During the earlier stage relaxation, the Li atoms near the Cu surface move downwards, while other Li atoms move upwards, which is depicted by the red arrows in Fig. 5(a). And then the deposited Li atoms gradually form a stable atomic nanofilm structure; the mean value of atom position of the final five hundred stable configurations of AIMD is chosen as representation for following calculations, which is illustrated in Fig. 5(a) and 5(b). We use "region-*n*" to represent a single region of deposited Li, for instance, the region-1 is the nearest Li atom distribution region to the Cu surface; the atoms in one region mean that their *z*-coordinates are close. We count the number of Li atoms in each region from the final two hundred steps in AIMD calculations, as shown in Fig. 5(c). After hundreds of AIMD calculation steps started from the initial configuration, the regional distribution structure becomes stable, there are occasionally Li atoms moving in the adjacent regions because of randomness, but the average number of atoms in each region remains basically unchanged. The Li atoms at the interface are quite stable except tiny vibrations, and the corresponding region-1 contains more Li atoms than other regions. This phenomenon agrees with the adsorption energy results from our static simulation in previous section. To further explore the interface effect, we also calculate the charge density difference (CDD, $\Delta\rho$) of the stable AIMD configuration, which is shown in Fig. 5(d). The CDD can be used to describe the interaction between the support and the adsorbate[40]. It is defined as:

$$\Delta\rho = \rho_{adsorbate/support} - \rho_{adsorbate} - \rho_{support}$$

where $\rho_{adsorbate/support}$, $\rho_{adsorbate}$ and $\rho_{support}$ are the corresponding charge densities of the combined system, the deposited adsorbate (Li) and the bare support (Cu), respectively. Fig. 5(d) illustrates that the charge rearrangement mainly occurs around the Li-Cu interface, which significantly enhances the bonding interactions between Li and Cu atoms at the interface. It indicates that the deposited Li atoms at interface suffer obvious adsorption effect from the Cu substrate. Besides, the distance between region-1 and Cu surface is 2.05 Å, and the distance between region-1 and region-2 is 2.89 Å, much larger than Li-Cu interface layer, which also indicates the strong interface effect.

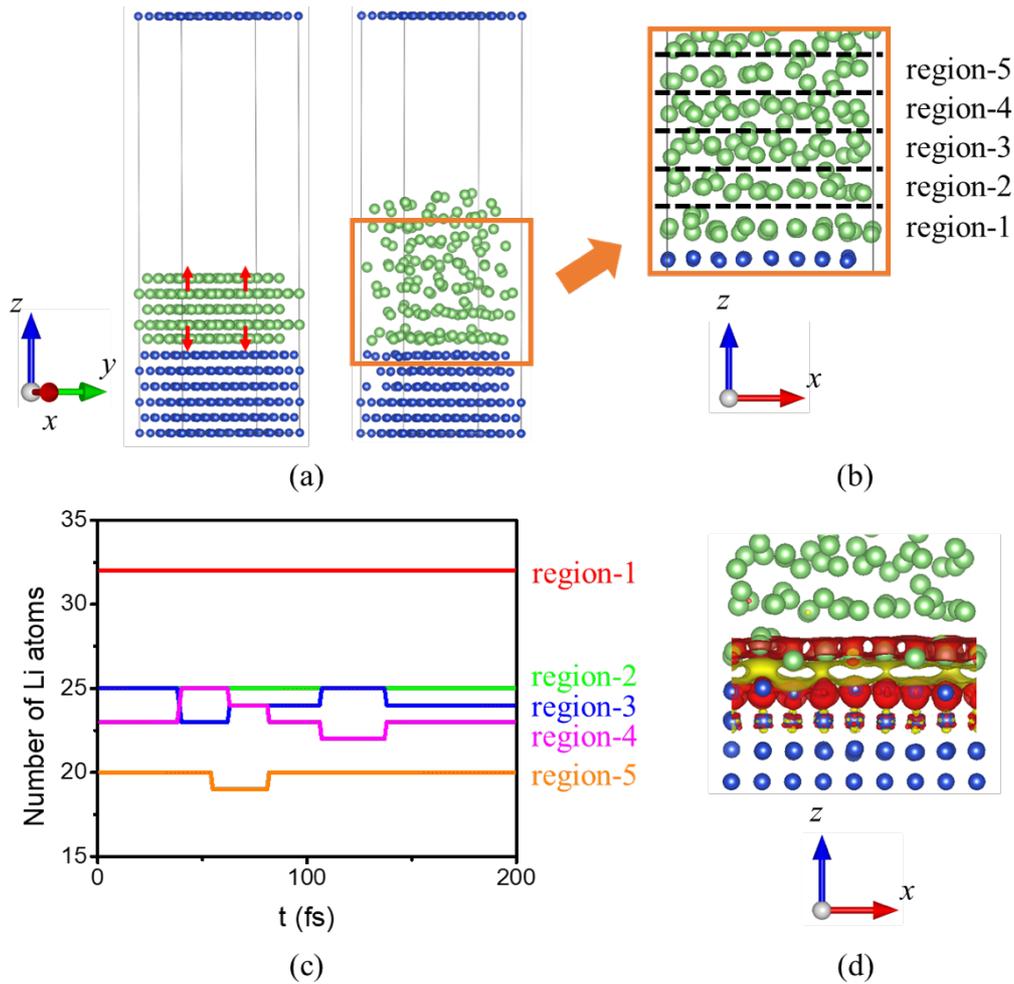

**Fig. 5** (a) The initial configuration and stable regional distribution configuration from AIMD calculation. The red arrows indicate the directions that the atoms move during the relaxation from initial structure. (b) The regional distribution structure of deposited Li atoms. The black dotted line represents the boundary between regions. (c) The number of Li atoms in each region for the final two hundred steps of AIMD calculations. (d) The charge density difference for stable AIMD configuration. The isosurface value is fixed at 0.001 e/bohr$^3$ with yellow and red colors, denoting the charge accumulation and charge depletion, respectively. The regions 1 to 5 are defined in the following way: the interface Li atoms bonded to Cu atoms are in region-1 with its position defined as the mean value of $z$-coordinates of these Li atoms. The Li atoms within a vertical distance of 3 Å from the region-1 position are in region-2 (excluding the atoms in region-1), with its position defined as mean value of the $z$-coordinates of these Li atoms. Other regions are defined in a similar way as region-2. The region boundary is located at the center of the region $n$ and region $n+1$ position.

According to previous studies[8,30-32], the dendrite problem in the deposition process is mainly caused by the in-plane compressive stress of the deposited film. In order to quantify the interaction between atoms in one region, which represents the in-plane effect, we first define the average atomic distance in one region. Considering the uniform distribution of atoms in an area, such as a square with its side as $l$ in Fig. 6(a), if all the distances of each pair of adjacent atoms are the same and the number of atoms in the area is $N=25$, the average atomic distance is $l/5$. By analogy, when the atoms are of random distribution, as shown in Fig. 6(a), if the number of atoms in the area is $N$, the average atomic distance could be defined by $l/\sqrt{N}$. Then we count the number of Li atoms of the stable AIMD configurations and calculate the corresponding average Li atomic distance of each region, which is presented in Fig. 6(b). From region-1 to region-5, the number of Li atoms within each region decreases; meanwhile, the corresponding average atomic distance becomes larger. The results are consistent with our conclusion in static calculations: lower adsorption energy tends to hold more deposited Li atoms aggregating near the Cu surface. To compare with unstressed pristine Li metal, we also draw dashed lines, representing the average distance value and corresponding number of atoms of Li metal (one atomic layer of the indices of crystal face (100)), in the calculation cell in Fig. 6(b). The in-plane Li-Li distances in all deposited regions are smaller than that of pristine Li metal, which indicates that the deposited Li atoms suffer in-plane compressive stress.

Interestingly, the compressive strain shows the largest magnitude (-25.4 %) at the Li-Cu interface, but it reduces dramatically to -5.5% with only several Li deposition regions (~2 nm), as shown in Fig. 6(c). It is expected that the strain will reduce to zero as more Li is deposited. Besides, Fig. 6(c) indicates that there is a giant strain gradient distribution (~$10^8$/m, comparable to the strain gradient in other systems[41-42]) from the Li-Cu interface to deposited Li surface, which makes the deposited atoms at the bottom tend to move upwards with fluctuations, causing the dendrite growth. It is interesting to show that the Li deposition induces the giant strain gradient and hence stress gradient. Compared to the compressive stress revealed in the previous investigations[29], the stress gradient enhances the tendency of Li surface instability and thus causes the dendrite growth during the Li deposition. We have recently found the "flexo-diffusion" effect, revealing that the strain gradient has a significant role in Li diffusion[43]. Here, we show that the strain gradient distribution formed during Li deposition may be the driving force of Li dendrite growth in Li-metal-based rechargeable batteries.

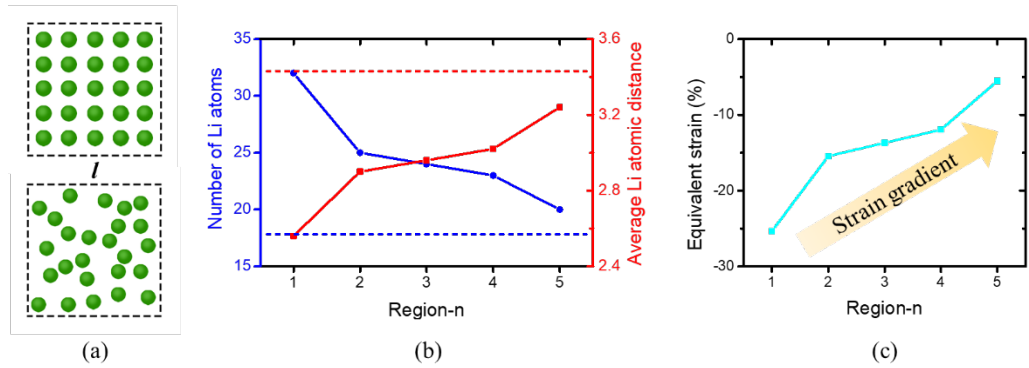

**Fig. 6** (a) Illustration of uniform distribution and random distribution of Li atoms. (b) The number of Li atoms (blue color) and average Li atomic distance (after projection to *xy*-plane, red color) in one region as a function of region number. The red dotted line represents the distance of Li atom in one atomic layer of Li metal (the indices of crystal face are (100)): 3.43 Å. Accordingly, the blue dotted line represents the number of Li atoms accommodated in the calculation cell for that atomic distance: 17.82. (c) The change of equivalent strain of different regions. The strain is defined as , where $a_0$ is the lattice constant of Li metal (3.43 Å), *a* is the average Li atomic distance of region, considered as the equivalent lattice constant.

According to the macroscopic theoretical research[32], the chemical potential at surface is higher than that at the grain boundary during growth due to supersaturation of adatoms on the surface. The increase in surface chemical potential makes atoms flow into the grain boundary, resulting in a compressive stress in the film. In fact, this phenomenon means that more Li atoms aggregate in close *z*-coordinates than pristine Li metal. Therefore, the deposited Li atoms are equivalent to suffering compressive stress in *xy*-plane. Based on our calculations of ground states of deposited Li atoms on

Cu substrate, the similar conclusion could be obtained. Owing to the lower adsorption energy and the stronger atomic interaction near the Cu surface, the deposited Li atoms will aggregate at Cu surface easily and therefore the deposited Li film suffers in-plane compressive stress. In addition to the compressive stress, we find there is huge strain gradient distribution decaying from Li-Cu interface to the surface, which makes the deposited atoms at the bottom tend to move upwards with fluctuations, causing the dendrite growth.

## 4 Conclusions

We investigate Li plating behaviors on Cu substrate and strain gradient driven Li dendrite growth mechanism at the atomic-scale using the first-principles calculations and *ab initio* molecular dynamics simulations. Our results show that the deposited Li atoms form a stable atomic nanofilm structure in the interface. Owing to the Li-Cu interface effect, the adsorption energy of Li atom near the Li-Cu interface is lower than those located near deposited surface, causing more deposited Li atoms to aggregate at the Li-Cu interface and hence induces compressive stress in the interface. There is also a giant strain gradient distribution from Li-Cu interface to the deposited Li surface. Therefore, the deposited atoms at the bottom tend to move upwards with perturbation due this strain gradient, which may cause the dendrite growth. Our results suggest that the in-plane compressive strain and strain gradient during Li deposition could be the driving force for Li dendrite growth from the perspective of atomic-scale. This provides a possible atomic-scale origin of the Li dendrite growth and will be useful for the optimization of Li-metal-based rechargeable batteries. More investigations, especially the experiment studies, are needed to verify this atomic origin of Li dendrite growth.